\documentclass[lettersize,journal]{IEEEtran}
\usepackage{amsmath,amsfonts}
\usepackage{algorithm}
\usepackage{array}
\usepackage[caption=false,font=normalsize,labelfont=sf,textfont=sf]{subfig}
\usepackage{textcomp}
\usepackage{stfloats}
\usepackage{url}
\usepackage{verbatim}
\usepackage{graphicx}
\usepackage{cite}
\usepackage{algpseudocode} 
\begin{document}

\title{JExplore: Design Space Exploration Tool for Nvidia Jetson Boards}


\author{
    \IEEEauthorblockN{Basar Kutukcu\IEEEauthorrefmark{1}, Sinan Xie\IEEEauthorrefmark{1}, Sabur Baidya\IEEEauthorrefmark{2}, Sujit Dey\IEEEauthorrefmark{1}} \\
    \IEEEauthorblockA{\IEEEauthorrefmark{1}Department of Electrical and Computer Engineering, University of California San Diego, CA, USA
    \\\{bktkc, six004, dey\}@ucsd.edu}\\
    \IEEEauthorblockA{\IEEEauthorrefmark{2}Department of Computer Science and Engineering, University of Louisville, KY, USA
    \\sabur.baidya@louisville.edu}
}



\maketitle

\begin{abstract}
Nvidia Jetson boards are powerful systems for executing artificial intelligence workloads in edge and mobile environments due to their effective GPU hardware and widely supported software stack. In addition to these benefits, Nvidia Jetson boards provide large configurability by giving the user the choice to modify many hardware parameters. This large space of configurability creates the need of searching the optimal configurations based on the user's requirements. In this work, we propose JExplore, a multi-board software and hardware design space exploration tool. JExplore can be integrated with any search tool, hence creating a common benchmarking ground for the search algorithms. Moreover, it accelerates the exploration of user application and Nvidia Jetson configurations for researchers and engineers by encapsulating host-client communication, configuration management, and metric measurement.
\end{abstract}


\section{Introduction}

The recent advancements of generative artificial intelligence have been affecting many different areas such as engineering, art, writing, chat for learning, etc. Many industries in these areas may benefit from running generative artificial intelligence models on private and mobile systems. Robotics, automotive, and consumer electronics are examples of such industries. There are many mobile hardware systems that can run generative artificial intelligence workloads such as mobile phones \cite{mobileLLM}, embedded systems with GPU support \cite{divcon} and FPGAs \cite{fpgallm}. However, these usually require lots of modifications to the models developed in common deep learning libraries such as PyTorch and HuggingFace.

Nvidia GPUs are dominant hardware for deep learning workloads due to their computation power and widely adopted software stack. They have the advantage of being the most common hardware backend for many deep learning tools. This usually translates to ease of using different systems as long as they both have capable Nvidia GPUs without leaving the development ecosystem. This creates the first advantage of the Nvidia Jetson boards. Any model developed in the common deep learning libraries is directly usable in Jetson boards as they have the same Nvidia GPU architectures as the other Nvidia GPUs. The second advantage is that, even though small in size, Jetson boards carry quite powerful GPUs. Considering large language models, the workhorse of generative artificial intelligence, require large computation powers, a powerful hardware backend is a requirement as much as it is an advantage.

Another advantage of Jetson boards is their configurability. Jetson boards allow users to change various hardware parameters such as number of CPU cores, CPU frequency, GPU frequency, EMC frequency etc. Modifying these parameters naturally results in changes in various performance metrics, such as application latency and power consumption. Finding close to optimal configurations in this vast search space requires intelligent search algorithms. These algorithms need to explore the search space by sampling configurations on the system. Since these explorations are hard to implement and take long time, many of the search algorithms are tested on exhaustively collected small search space \cite{pal} or synthetic problems \cite{DTLZ}. However, using a small search space cannot test the scalability of the search algorithm and using a synthetic problem cannot test the algorithms reliability in real applications. Therefore, there's a need for large, real world application based search spaces and a tool to interface this search space and search algorithms.

Based on the advantages of Jetson boards and the necessity of an exploration tool for the search algorithms, we propose JExplore. The main contributions of this tool and the paper are:

\begin{itemize}
    \item Encapsulating communication, configuration management, metric measurement and result saving into simple and yet effective APIs
    \item It can communicate with multiple Jetson boards, allowing batch sampling algorithms to work faster
    \item Creating a common benchmarking ground for the search algorithms for testing in a large and real world application based search space
    \item We have published the code as open source at https://github.com/basarkutukcu/JExplore
\end{itemize}

\section{Motivation and Related Work}

Improvements in technology and a large variety of requirements from different applications result in large design spaces in many different platforms. As a result, search algorithms become important in identifying optimal solutions in these large design spaces. Many different search algorithms are proposed to search these design spaces. In \cite{ehvi}, Bayesian optimization with expected hypervolume improvement based acquisition function is used. In \cite{NSGA_II}, an evolutionary algorithm is used to solve large search spaces. In \cite{divcon, sherlock}, customized Bayesian optimization algorithms are used to search design spaces. While there are many algorithms like these, there are not many opportunities to compare these algorithms. One common methodology is to use synthetic problems such as \cite{DTLZ} to benchmark and compare algorithms. While synthetic problems can create lots of different conditions and are very helpful to quickly benchmark a synthetic problem, they are not the real target of these search algorithms. Search algorithms are developed to solve real world problems and therefore testing against real world problems are important.

There are also many different real world problems that create design spaces. In \cite{sherlock}, FPGA design problems are used to benchmark the search algorithm. In \cite{pal}, an FPGA design space, an ASIC design space and a compiler optimization design space are used. In \cite{divcon}, combined search spaces of Nvidia Jetson hardware and various kinds of software parameters are used. In \cite{ansor}, compiler optimization search space is used.

The improvements in generative artificial intelligence are promising in terms of the future of generative AI based applications. A large portion of these applications can be on embedded systems to extend their benefits deep into people's lives since embedded systems are ubiquitous in life. Since Nvidia GPUs are one of best hardware that can execute these workloads, exploring the design spaces posed by Nvidia Jetson is an important research problem. Therefore, our tool can accelerate the development of the search algorithms and provide a common useful benchmarking ground for the search algorithms. 

\section{JExplore}

\begin{figure*}
    \centering
    \includegraphics[width=1\linewidth]{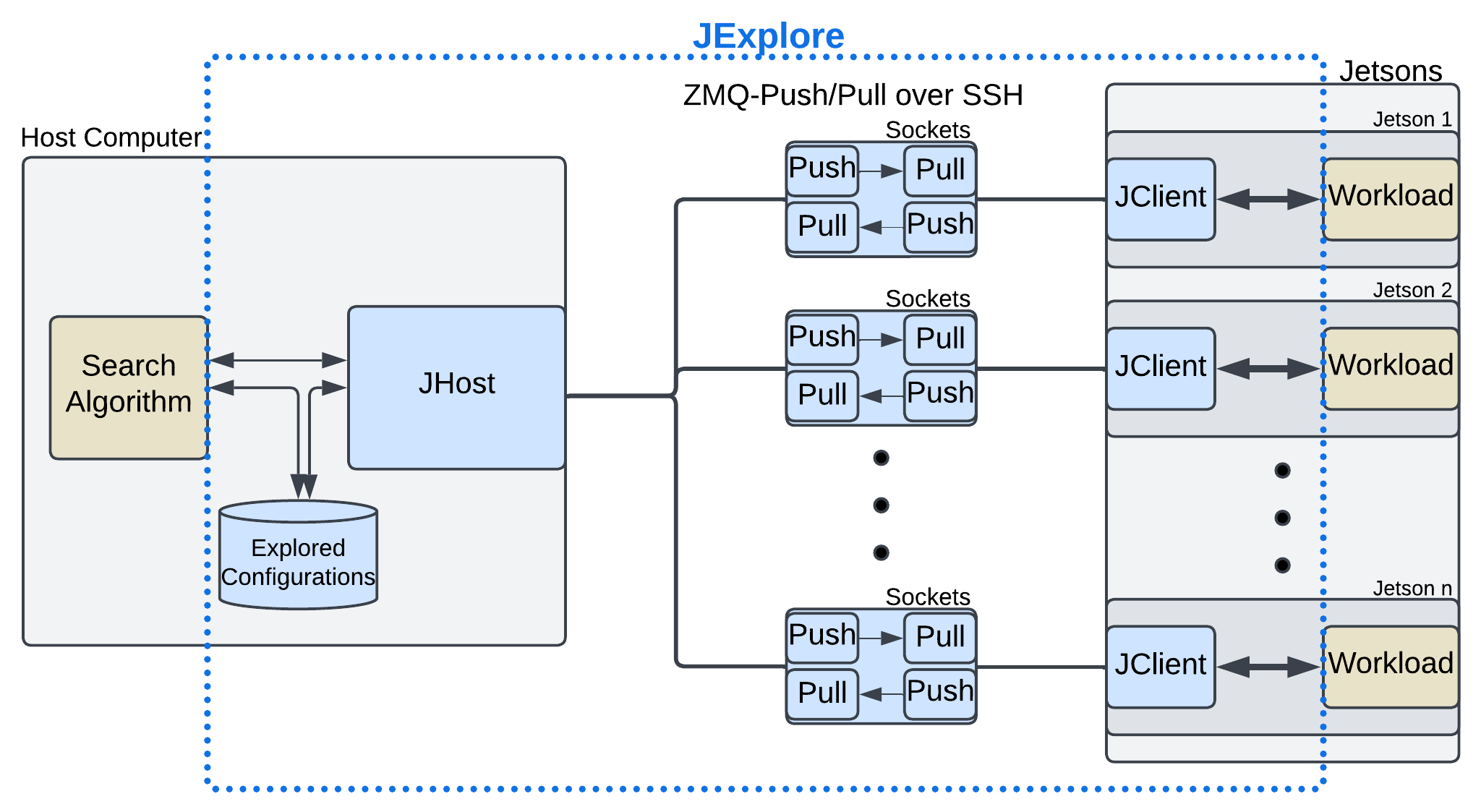}
    \caption{JExplore overview}
    \label{fig:overview}
\end{figure*}

The overall architecture of JExplore is shown in Figure \ref{fig:overview}. JExplore consists of two main components which are JHost and JClient. JHost runs on the host computer and works as an interface between the user defined search algorithm and Jetson boards. It also has utility functions such as saving the explored search space in CSV format. JHost and JClients on Jetson boards communicate using ZMQ Push/Pull sockets over SSH. IP addresses and ports are defined by the user when the instances of these classes are initialized. 

JClient runs on the Jetson device. It has 3 main capabilities. The first one is to configure the Jetson device and the application based on the input it receives from JHost. There are many modifiable hardware parameters in the Jetson systems, which make them flexible embedded systems. The full list of these parameters, the number of values they can take and the ranges of these values are shown in Table \ref{tab:hw_feat}. JClient can configure these parameters using its subcomponent called JConfig. Nvidia normally gives power modes to the user to modify these parameters. However, there are usually 5-10 modes provided by Nvidia. On the other hand, JConfig can explore the full ranges of parameters which results in much more fine-grained search space. The second capability of JClient is the measurement using the abstract class JMeasure. It is designed as abstract class to allow the users of JExplore to define their own custom measurement functions if necessary. However, JExplore includes some fundamental measurement functions. Time, power consumption and memory consumption are measured by the JTime, JPower, and JMemory subclasses. Each measurement can be enabled/disabled when the JClient instance is defined. The last capability is the communication with the host computer. This is done through ZMQ push and pull sockets that are implemented over SSH. ZMQ sockets define a communication protocol where each socket has a certain job. The SSH is used to remove the requirement of host computer and jetson devices to be on the same subnet while keeping the connection secure. The details of the connection related functions are hidden in JHost and JClient classes, allowing users of JExplore to use the communication system without much effort.

\begin{table}[!tb]
\centering
\begin{tabular}{|c|c|}
 \hline
 Parameter name & \# of possible values (range) \\
 \hline
 \hline
 \# of CPU cores - Cluster 1 & 4 (1-4)
 \\ 
 \hline
 \# of CPU cores - Cluster 2 & 5 (0-4)
 \\
 \hline
 \# of CPU cores - Cluster 3 & 5 (0-4)
 \\
 \hline
 Cluster 1 CPU Frequency & 29 (115MHz - 2.2GHz)
  \\
\hline
 Cluster 2 CPU Frequency & 29 (115MHz - 2.2GHz)
  \\
\hline
 Cluster 3 CPU Frequency & 29 (115MHz - 2.2GHz)
  \\
 \hline
 GPU Frequency & 11 (306MHz - 1.3GHz)
 \\
 \hline
 EMC Frequency & 4 (204MHz - 3.2GHz)
 \\
\hline
\end{tabular}
\vspace{1mm}
\caption{Nvidia Jetson Orin Modifiable Hardware Parameters}
\label{tab:hw_feat}
\end{table}

The workload should be implemented on Jetson side and used by JClient. An example pseudocode of a simple system with JHost and one JClient is shown in Algorithm \ref{alg:code}. JClient receives the configurations which are suggested by the search algorithm. Then JClient configures the board and the software parameters of the workload. After that workload is run and the measurement results are sent to JHost. The workloads can be anything as JExplore is agnostic to the workload. 

\begin{algorithm}[t]
	\caption{Simple example of JHost and JClient} 
	\begin{algorithmic}[1]
            \Procedure{JHost}{$testConfigs$}
                \While {$testConfigs$ are available}
                    \State Push $testConfig$ to JClient
                    \State Pull $results$ from JClient
                \EndWhile
            \EndProcedure
            \Procedure{JClient}{}
                \While {$testConfigs$ are available}
                    \State Pull $testConfig$ from JHost
                    \State Configure Jetson using $testConfig$
                    \State Configure workload using $testConfig$
                    \State Run workload
                    \State Push $testConfig$ to JHost
                \EndWhile
            \EndProcedure
	\end{algorithmic} 
\label{alg:code}
\end{algorithm}

\section{Experiments}

We have implemented 2 generative artificial intelligence workloads to test and demonstrate the JExplore capabilities. The first workload is text generation using input text prompt by a large language model. The second workload is text generation using image and text input prompt by a large language and vision assistant model.

\subsection{Llama2-7B}

\begin{figure}
    \centering
    \includegraphics[width=1\linewidth]{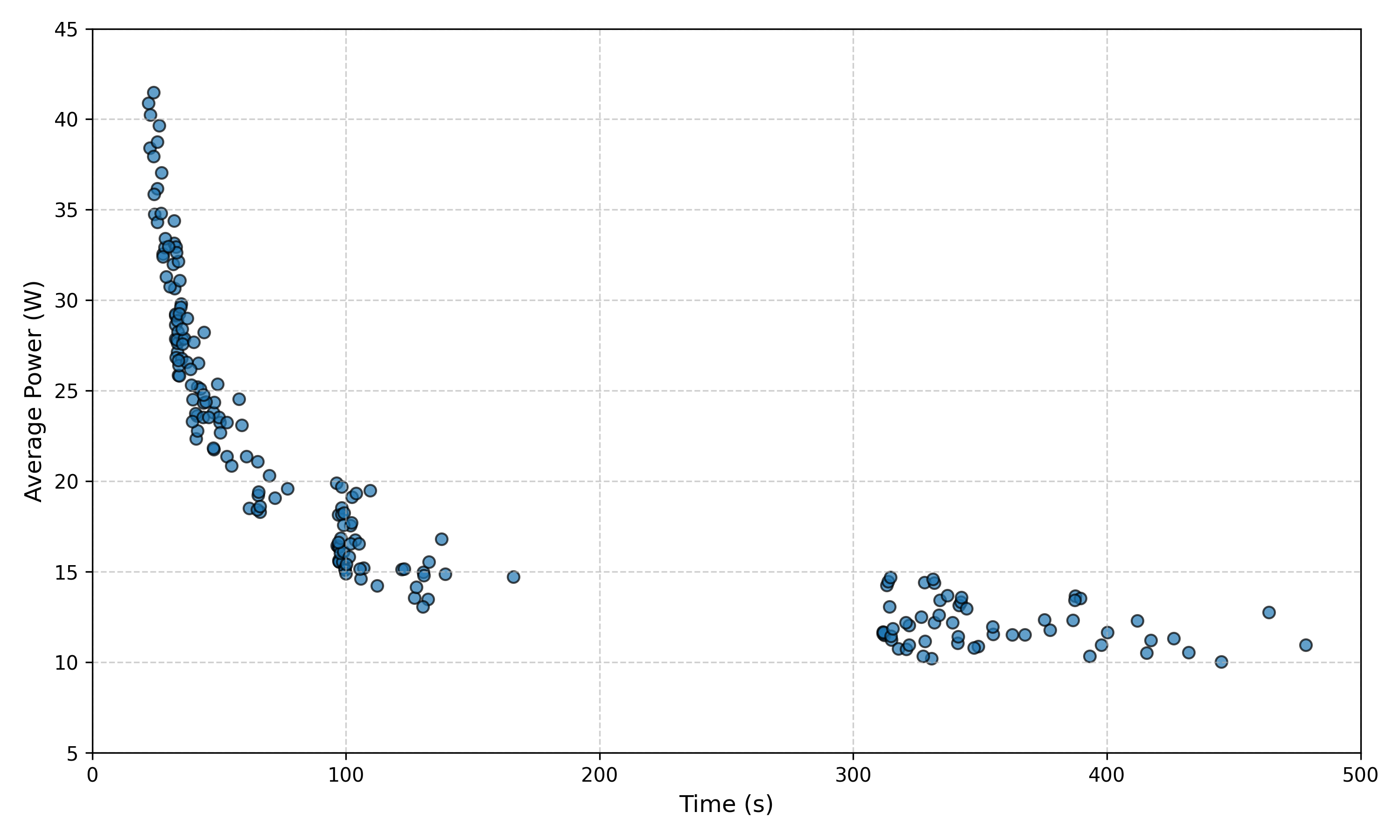}
    \caption{Power and time measurements of 200 different Jetson Orin configurations by JExplore for running Llama2-7B}
    \label{fig:llama}
\end{figure}

We have used the Llama2-7B \cite{llama} model as workload. It is a large language model that takes an input text prompt and generates text based on the prompt. We have provided the following text as prompt: "\textit{Can you explain the importance of renewable energy and how it can help combat climate change? Please explain in 150 words.}". We randomly sampled 200 Nvidia Jetson Orin configurations from the search space shown in Table \ref{tab:hw_feat}. We have used greedy sampling for token generation so that all inferences generate the same output. The results for the average power consumption and inference time of Llama2-7B running on 200 configurations are shown in Figure \ref{fig:llama}. We observe that power consumption and inference latency are inversely correlated as expected. As a result, a clear pareto frontier emerged in the plot. The power ranges from 10W to 42W. The time is also taking values from ~20s to almost 500s.

\subsection{LLaVA-v1.5-7B}

\begin{figure}
    \centering
    \includegraphics[width=0.5\linewidth]{figures/cats.jpg}
    \caption{Input image to LLaVA workload}
    \label{fig:cats}
\end{figure}

\begin{figure}[t]
    \centering
    \includegraphics[width=1\linewidth]{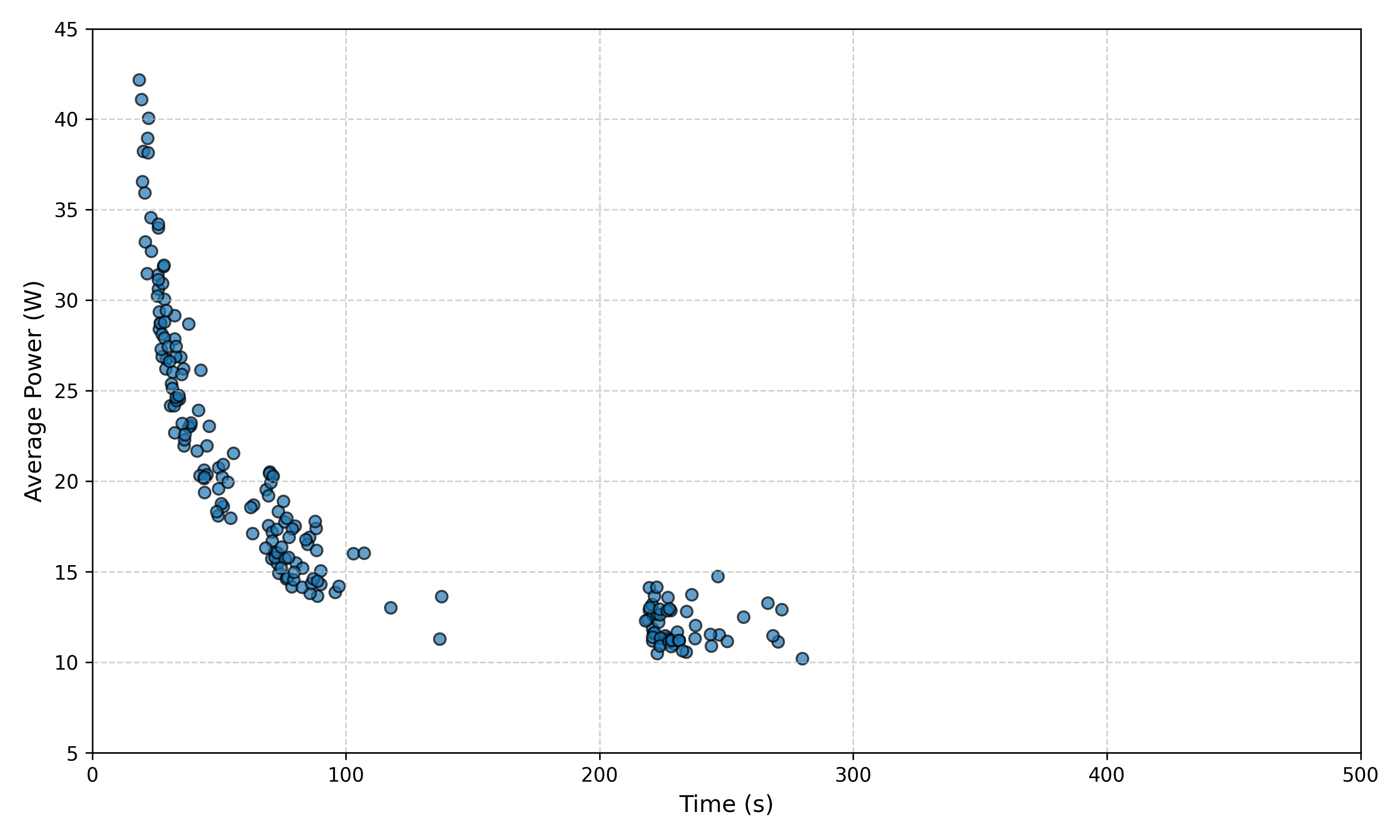}
    \caption{Power and time measurements of 200 different Jetson Orin configurations by JExplore for running LLaVA-v1.5-7B}
    \label{fig:llava}
\end{figure}

We have used LLaVA-1.5-7B \cite{llava} model as workload. It is a large multimodal model that takes an input image and text prompt and generates text based on the inputs. We have provided the image shown in Figure \ref{fig:cats} and the text prompt of "\textit{Tell me a bedtime story within 150 words based on this image}". Similar to the previous experiment, we have randomly generated 200 configurations and used JExplore to collect results. The results for average power consumption and the total latency are shown in Figure \ref{fig:llava}. The shape of the data is similar to the Llama experiment since we have used power and time metrics again.

There are some differences and some similarities between the two workloads. The power ranges between two workload are similar as both workloads can fully utilize the board to achieve maximum power consumption and have similar profiles to achieve minimum power consumption. However, the inference latency is different as LLaVA requires less time to run. This results in LLaVA having data points that are scattered more densely than Llama.

Another important common property that we can notice in both plots is the separate clusters from the main data points. In Figure \ref{fig:llama}, Llama has a separate cluster of points at time range 300s-500s. Similarly, there's a  cluster for LLaVA at time range 200s-300s as shown in Figure \ref{fig:llava}. Even though the exact range they show up is different, their patterns are similar in a sense that they are separate from the other points in the plots. Our analysis found that the reason these clusters show up is the EMC frequency. All the data points in these cluster for both workloads use the lowest EMC frequency. So the EMC frequency, unlike other hardware parameters, has a cut-off effect in time. In other words, decreasing the EMC frequency to the lowest value increases the inference time more than other hardware parameter changes. As a result, we see a jump in time for the data points scatter plot. JExplore makes this kind of analyses easy thanks to the ready to use utilities and benchmarking capabilities.

\section{Conclusion}

In this paper, we have proposed a benchmarking and design space exploration tool for Nvidia Jetson devices. Our framework is designed to be easily used by other researchers and engineers. It can accelerate search algorithm development by providing a quick testing framework. Moreover, it can provide common benchmarking ground for comparing search algorithms. We have implemented two generative artificial intelligence based workloads to demonstrate the capabilities of JExplore.

\section*{Acknowledgments}

\bibliographystyle{IEEEtran}
\bibliography{jexplore}

\end{document}